\documentclass[12pt]{iopart}

\usepackage{graphicx,bm}
\def\v#1{\mbox{\boldmath$#1$}}

\def\simgeq{\mbox{\raisebox{-1.0ex}{$\stackrel{>}{\sim}$}}}
\newcommand{\aver}[1]{\left\langle {#1} \right\rangle}
\newcommand{\ket}[1]{\left\vert {#1} \right\rangle}
\begin{document}

\title{Ferrimagnetic States in $S=1/2$ Frustrated Heisenberg Chains with Period 3 Exchange Modulation 
}

\author{K Hida}

\address{Divison of Material Science, Graduate School of Science and Engineering, Saitama University, Saitama, Saitama, 338-8570, JAPAN 
}
\ead{hida@phy.saitama-u.ac.jp}
\begin{abstract}
Ground state properties of the $S=1/2$ frustrated Heisenberg chain with period 3 exchange modulation are investigated  using the numerical diagonalization and DMRG method. It is known that this model has a magnetization plateau at $1/3$ of the saturation magnetization $M_{\rm s}$. On the other hand, the ground state is ferrimagnetic even in the absence of frustration if one of the nearest neighbour bond is ferromagnetic and others are antiferromagnetic. In the present work, we show that this ferrimagnetic state continues to the region in which all bonds are antiferromagnetic if frustration is strong.  This state further continues to the above mentioned $1/3$-plateau state. In between, we also find the noncollinear ferrimagnetic phase in which the spontaneous magnetization is finite but less than $M_{\rm s}/3$. The intuitive interpretation for the phase diagram is given and the physical properties of these phases are discussed. 

\end{abstract}

\pacs{75.10.Jm, 75.10.Pq}

\section{Introduction}
Frustrated quantum spin chains have been the subject of extensive studies for decades. One of the most remarkable phenomena driven by frustration is the transition to spontaneously dimerized ground state as demonstrated by the exact solution of Majumdar and Ghosh\cite{mg}. In the magnetic field, another type of translational symmetry breakdown is recently found by Okunishi and Tonegawa\cite{oku1,oku2} and Tonegawa {\it et al.}\cite{tone} resulting in a nontrivial magnetization plateau at one third of full magnetization $M_{\rm s}$. The present author and Affleck\cite{ha} have investigated the effect of period three exchange modulation on this plateau state. It turned out that the transition between the classical plateau state and the quantum plateau state takes place within the $1/3$ plateau state 

Another remarkable effect of frustration is the stabilization of ferrimagnetic ground state. Yoshikawa and Miyshita\cite{ym} proposed a model of frustrated quantum chain which shows a ferrimagnetic ground state. Remarkably, this model not only has a Lieb-Mattis type ground state in which  magnetization is fixed to a value determined by the difference of  number of sites of two sublattices but also has a noncollinear ferrimagnetic ground state where  magnetization is not a simple fraction of full magnetization. In this state, the local magnetization profile has an incommensurate structure.

In the present paper, we show that  the $S=1/2$ frustrated Heisenberg chain with period 3 exchange modulation  also shows  Lieb-Mattis type and noncollinear type ferrimagnetism in appropriate parameter range. The former continues to the $1/3$ plateau state investgated in ref. \cite{ha}.

This paper is organized as follows. In the next section, we present the model Hamiltonian. The ground state phase diagram obtained by numerical diagonalization is presented in section 3. The property of each phase is also discussed based on the density matrix renormalization group (DMRG) calculation. In the last section we summarize our results.

\section{Hamiltonian}
The Hamiltonian of the $S=1/2$ frustrated Heisenberg chain with period 3 exchange modulation is given by,
\begin{eqnarray}
{\cal H} &=&J\sum_{l=1}^{N/3} \left[(1-\alpha)\left(\v{S}_{3l-1}\v{S}_{3l} +\v{S}_{3l}\v{S}_{3l+1}\right)\right.\nonumber\\
&+&\left.(1+\alpha)\v{S}_{3l+1}\v{S}_{3l+2}\right]+J\delta\sum_{i=1}^{N}\v{S}_{i}\v{S}_{i+2}.
\label{ham}
\end{eqnarray}
where $\v{S_i}$ is the spin $1/2$ operator and $N$ is the number of sites. This model has a magnetization plateau at $1/3$ of the saturation magnetization $M_{\rm s}$ as investigated in \cite{ha}. On the other hand, it is obvious that the ground state is ferrimagnetic for $\alpha < -1$ even in the absence of frustration. In this paper, we concentrate on the ground state properties of this model in the region $-1 < \alpha <0$ and $\delta >0$.

\section{Ground State Properties}

\begin{figure}
\centerline{\includegraphics[width=100mm]{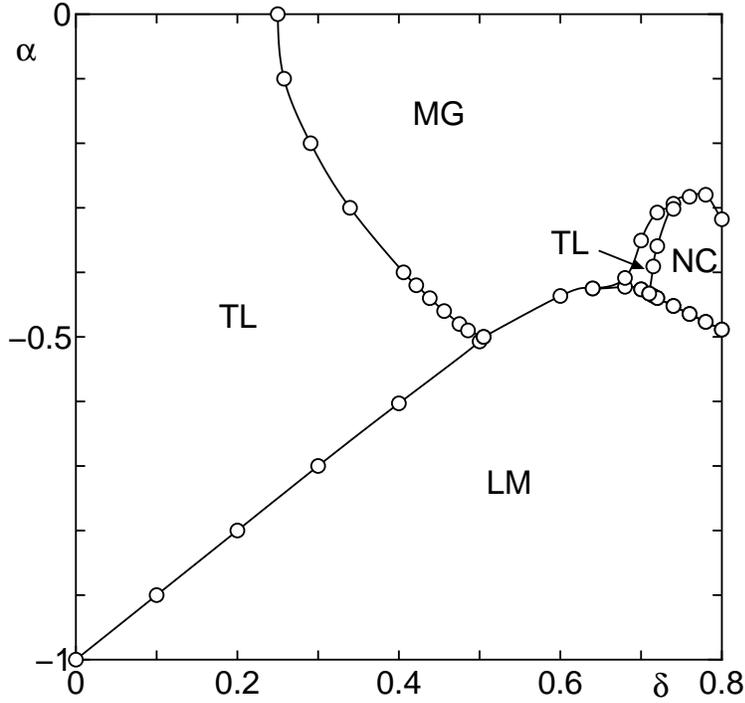}}
\caption{Ground state phase diagram for $-1 \leq \alpha \leq 0$ for $0 \leq \delta \leq 0.8$. TL, MG, NC and LM stand for Tomonaga-Luttinger liquid phase, Majumdar-Ghosh type dimer phase, noncollinear and Lieb-Mattis ferrimagnetic phases, respectively. }
\label{phase}
\end{figure}

The ground state phase diagram is obtained for $\alpha \leq 0$ and $0 \leq \delta \leq 0.8$ as shown in Fig. \ref{phase} by the numerical diagonalization of finie size system with $N=12,18$ and 24.  For $\delta >0.8$, the strong finite size effect prevents the precise deterimination of phase boundary.

For small $|\alpha|$, the ground state is the gapless Tomonaga-Luttinger liquid for small $\delta$ and the Majumdar-Ghosh type spontaneously dimerized phase  for larger $\delta$. The transition between these two phases is the Brezinskii-Kosterilitz-Thouless type transition and the phase boundary can be determined by the level spectroscpic method\cite{no} usung the numerical diagonalization data for $N=12, 18$ and 24.

\begin{figure}
\centerline{\includegraphics[width=80mm]{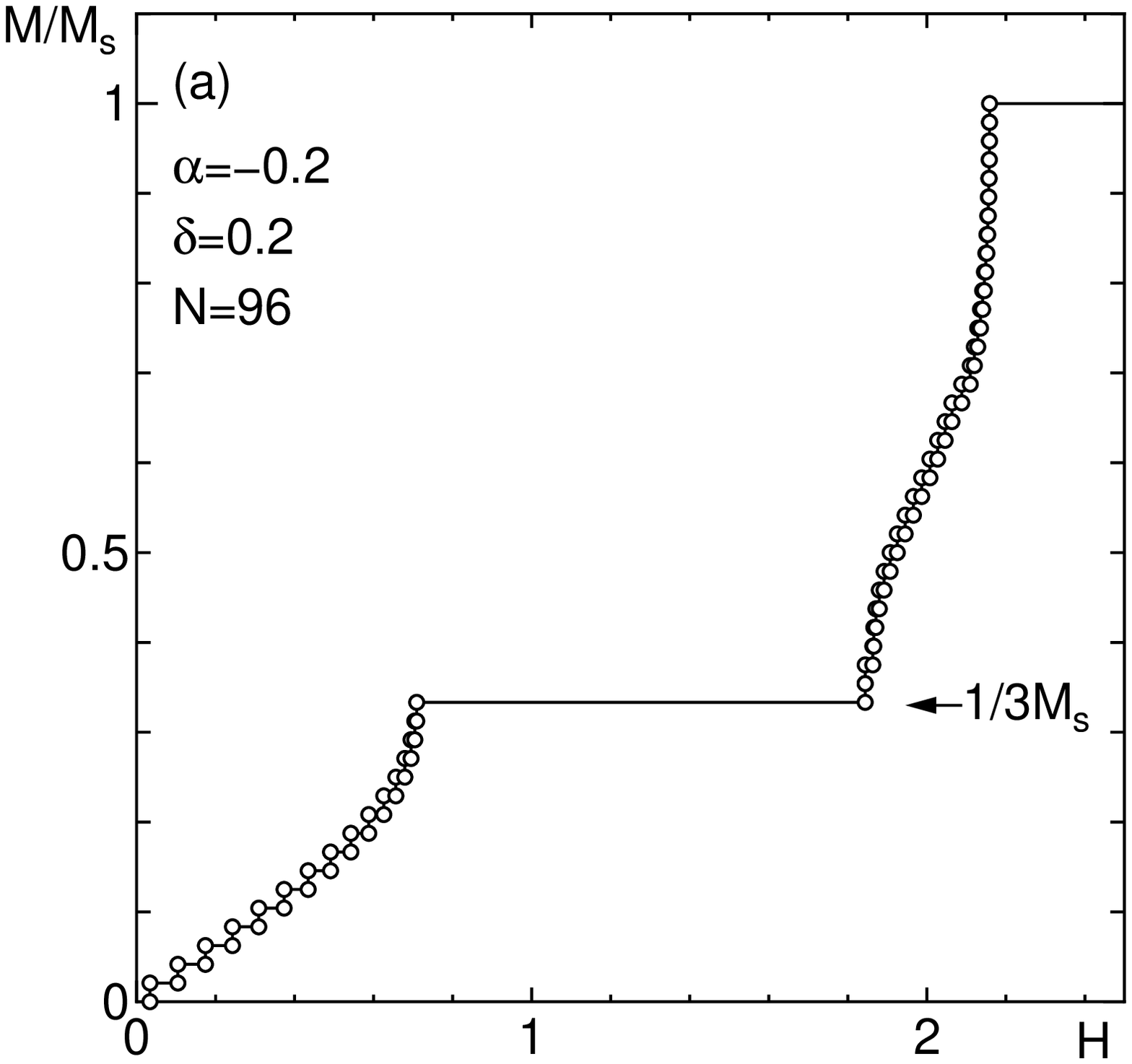}\includegraphics[width=80mm]{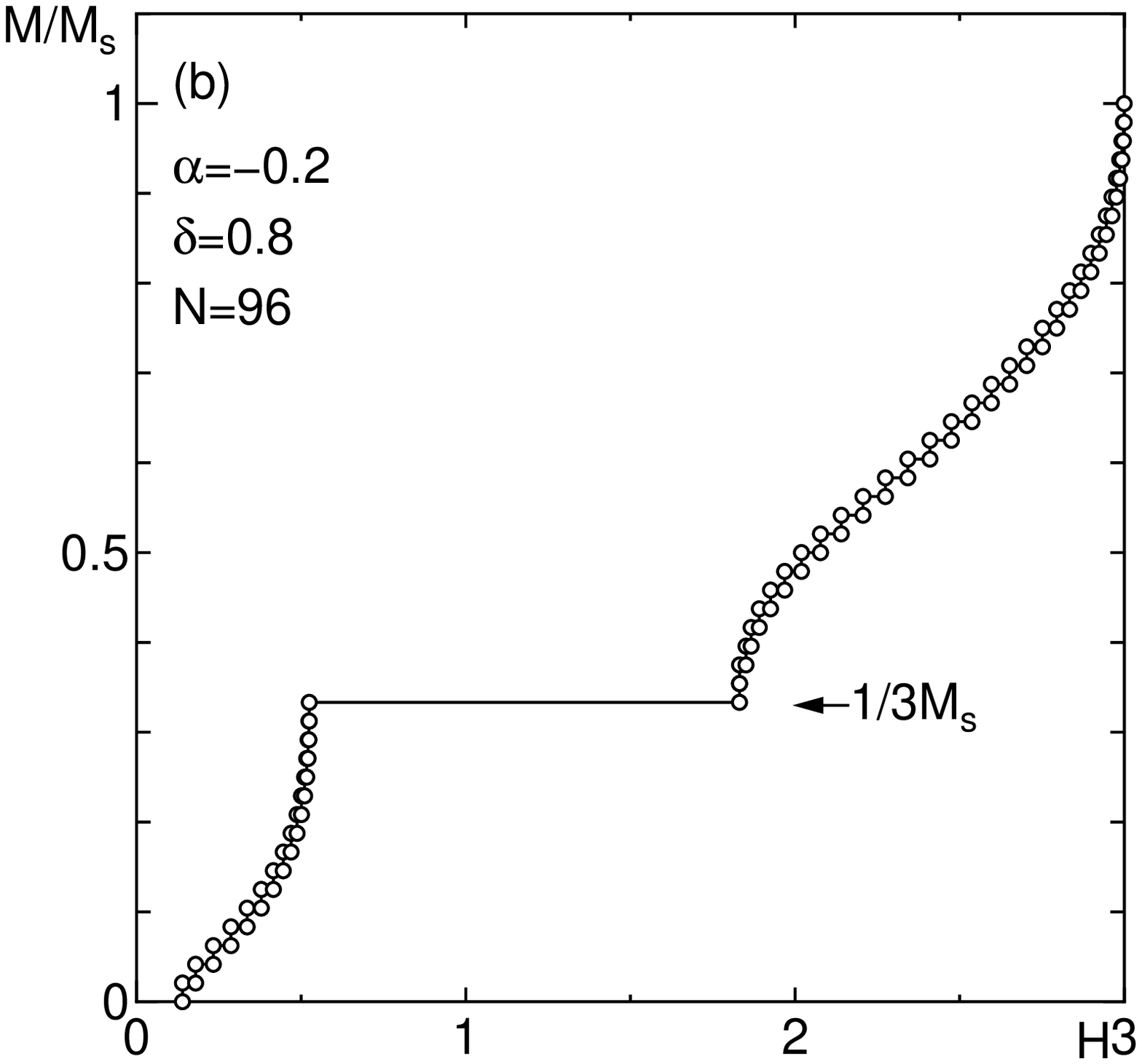}}
\caption{Magnetization curve in (a) Tomonaga-Luttinger liquid phase with $(\delta,\alpha)=(0.2,-0.2)$ and (b) Majumdar-Ghosh type dimer phase with $(\delta,\alpha)=(0.8,-0.2)$ for $N=96$ calculated by DMRG method. }
\label{magnm}
\end{figure}

Typical magnetization curves calculated by the DMRG method in these two nonmagnetic phases are shown in Fig. \ref{magnm} (a) for $(\alpha,\delta)=(-0.2,0.2)$  and (b) for $(\alpha,\delta)=(-0.8,0.8)$ and $N=96$ with open boundary condition. It is evident that the  spin gap is absent in the former case while it is present in the latter case. The magnetization plateau at $M=M_{\rm s}/3$ is always present reflecting the period 3 exchange modulation.

\begin{figure}
\centerline{\includegraphics[width=80mm]{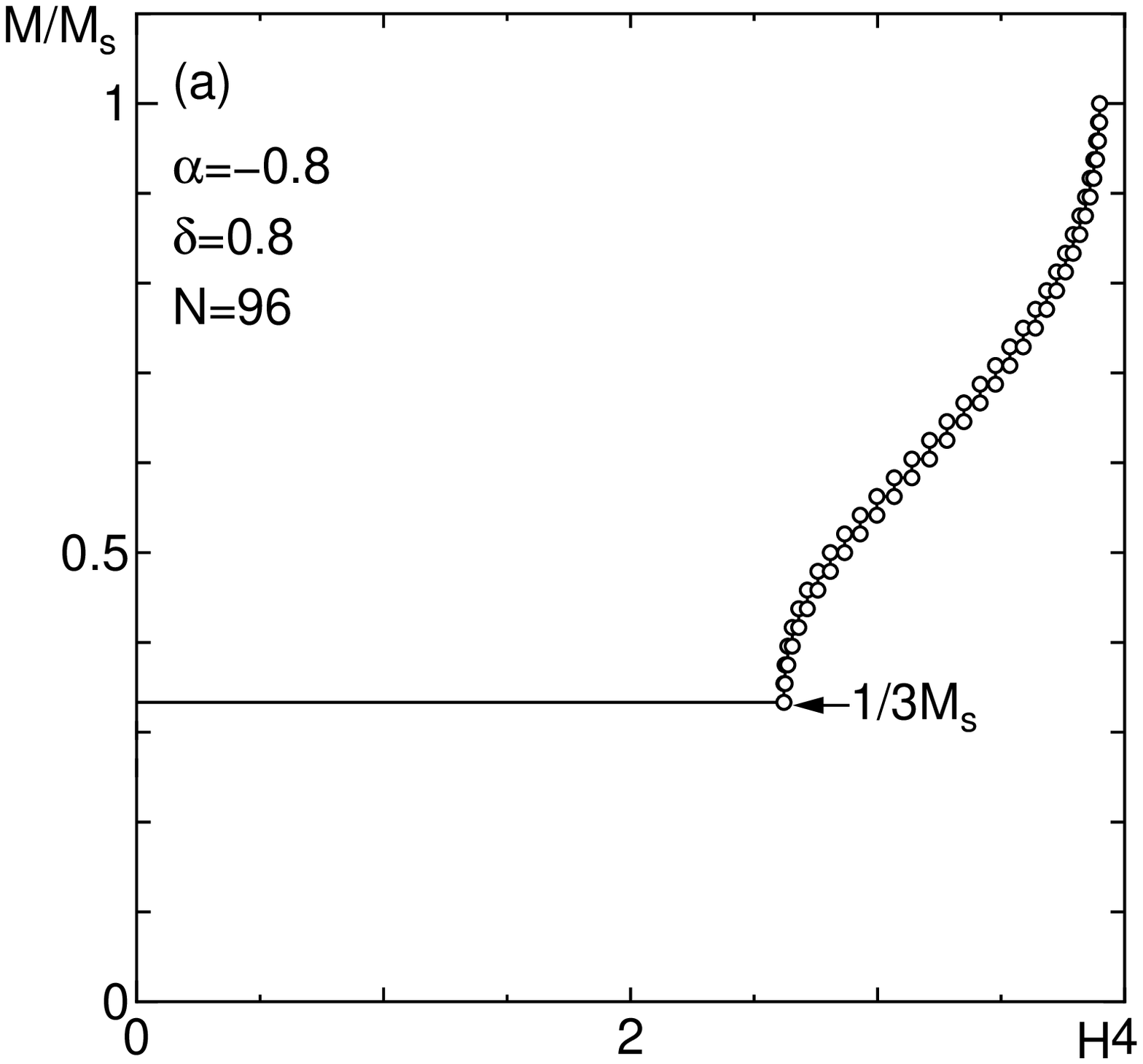}\includegraphics[width=80mm]{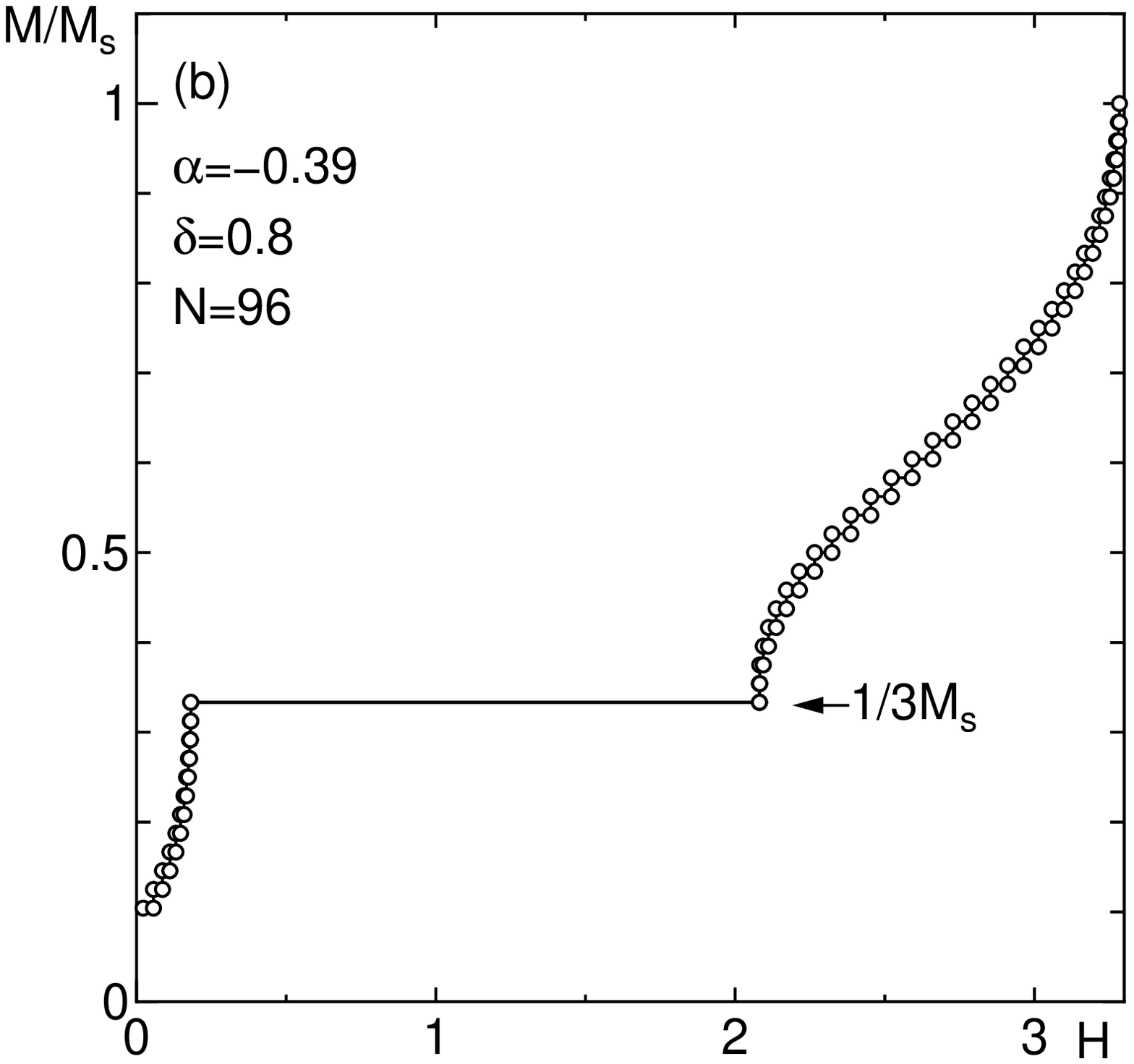}}
\caption{Magnetization curves in (a) Lieb-Mattis type ferrimagnetic phase with $(\delta,\alpha)=(0.8,-0.8)$ and (b) noncollinear ferrimagnetic phase with $(\delta,\alpha)=(0.8,-0.39)$ for $N=96$ calculated by DMRG method.}
\label{magmag}
\end{figure}

\begin{figure}
\centerline{\includegraphics[width=80mm]{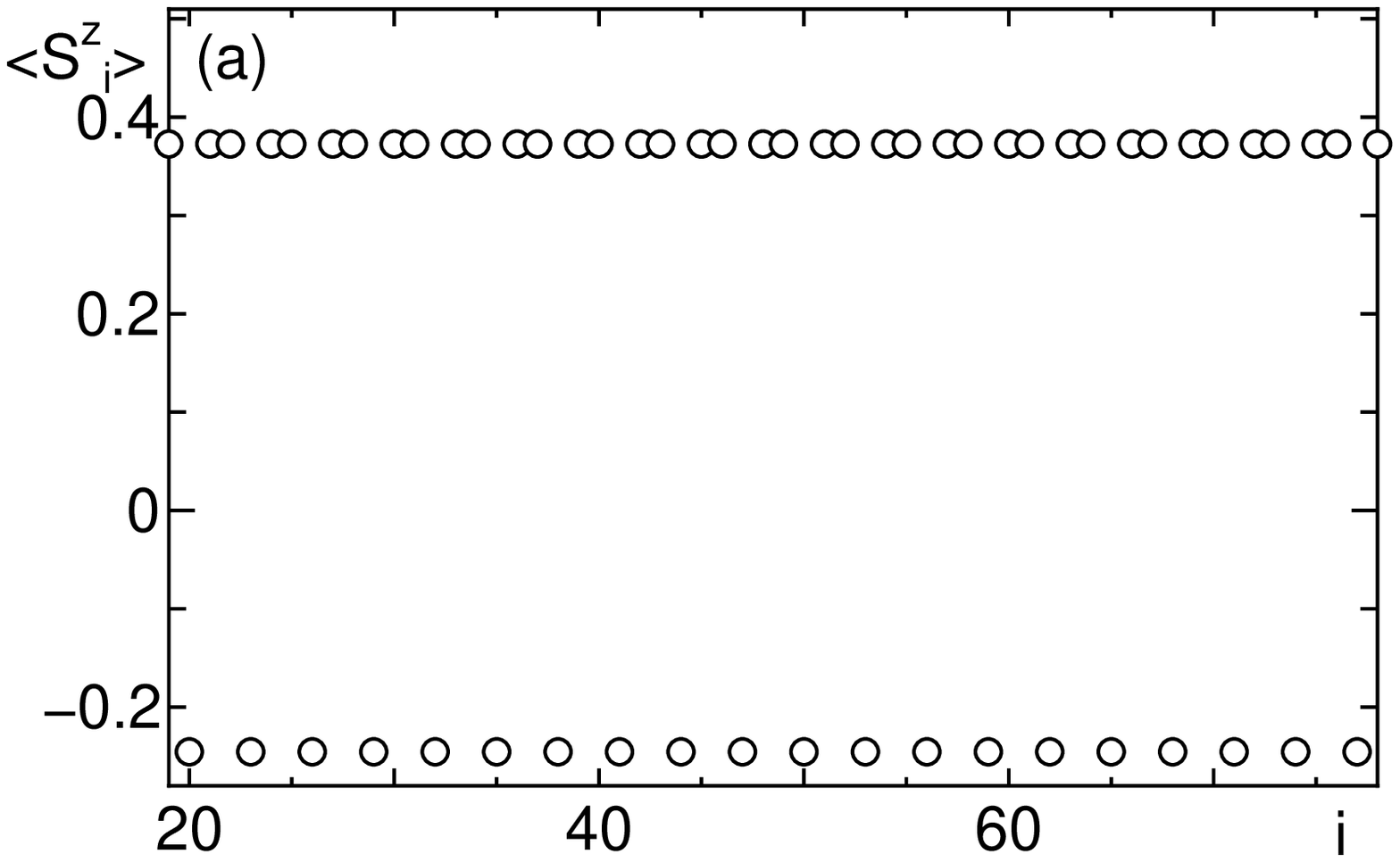}\includegraphics[width=80mm]{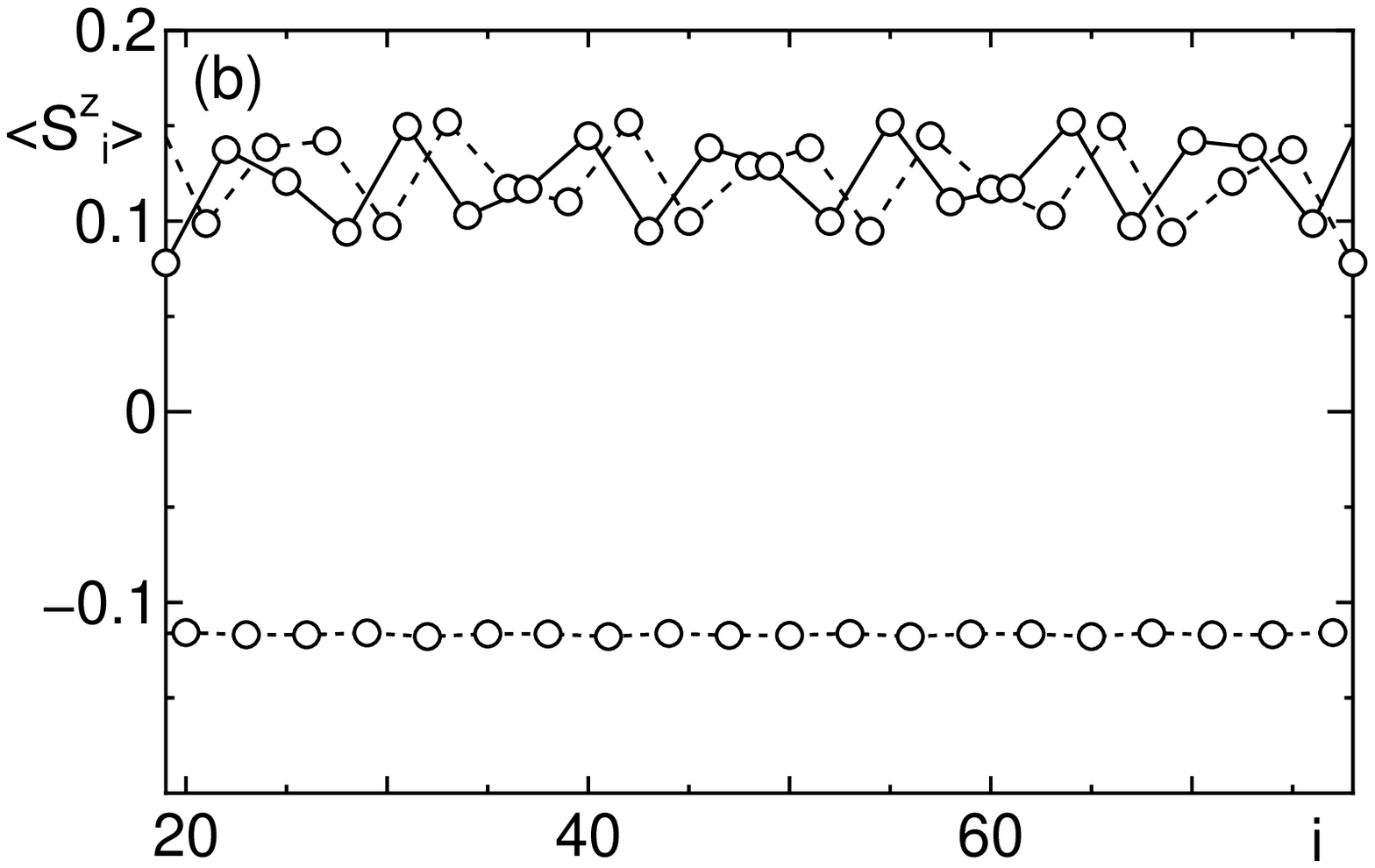}}
\caption{Local magnetization profile (a) in the Lieb-Mattis type ferrimagnetic phase with $\delta=0.7$ and $\alpha=-0.7$  and (b) in the noncollinear ferrimagnetic phase with $\delta=0.8$ and $\alpha=-0.39$  for  $N=96$ calculated by DMRG method.  Only the sites in the middle of the system $19 \leq i \leq 78$ is shown to exclude the  boundary effects.The lines are guides for eye drawn to make clear the incommensurate modulation of the magnetization profile.}
\label{lm}
\end{figure}

For large negative $\alpha$ and large $\delta$, the ground state is ferrimagnetic  with magnetization $M=M_{\rm s}/3$. Even though the present model is not strictly bipartite due to next nearest neighbour exchange interaction, this phase can be regarded as the Lieb-Mattis type ferrimagnetic phase because it is directly connected to the Lieb-Mattis ferrimagnetic state with $M=M_{\rm s}/3$ at $\alpha < -1$ and $\delta=0$. For small $\delta$ and $\alpha \simeq -1$, the three spins connected by the $(1-\alpha)$-bonds form an effective spin 1/2 doublet.

\begin{eqnarray}
\ket{\Uparrow}&=&\frac{1}{\sqrt{6}}\left(\ket{\uparrow\uparrow\downarrow}-2\ket{\uparrow\downarrow\uparrow}+\ket{\downarrow\uparrow\uparrow}\right)\\
\ket{\Downarrow}&=&\frac{1}{\sqrt{6}}\left(\ket{\downarrow\downarrow\uparrow}-2\ket{\downarrow\uparrow\downarrow}+\ket{\uparrow\downarrow\downarrow}\right)
\end{eqnarray}
 By elementary manipulation, the effective exchange interaction between these effective spins turns out to be $2J(1+\alpha-\delta)/9$. Therefore the ground state in this region is a Lieb-Mattis type ferrimagnetic state for $1+\alpha < \delta$ and the Tomonaga-Luttinger liquid state otherwise. Typical magnetization curve calculated by DMRG method in this phase is shown in Fig. \ref{magmag}(a) for $(\alpha,\delta)=(-0.8,0.8)$ and $N=96$ with open boundary condition. Comparing this magnetization curve with Fig. \ref{magnm}, it is clear that this ferrimagnetic state continues to the {\it classical} plateau state in the nonmagnetic phases. It should be noted that the {\it quantum} plateau state is realized only for $\alpha >0$\cite{ha}. The local magnetization profile $\aver{S^z_i}$ calculated by the DMRG method clearly shows 3 sublattice structure as shown in Fig. \ref{lm}(a) for $(\alpha, \delta)=(-0.7, 0.7)$ and $N=96$ with open boundary condition.  It should be also noted that spins are not fully polarized even in the Lieb-Mattis type phase although the total magnetization is exactly quantized to $M_{\rm s}/3$.

With the decrease in $|\alpha|$, the ferrimagnetic state with magnetization less than $M_{\rm s}/3$ appears for $\alpha \simgeq 0.72$. This phase has similarity with the noncollinear ferrimagnetism studied by Yoshikawa and Miyashita\cite{ym}. As a representative, the magnetization curve in this state calculated by the DMRG method is presented in Fig. \ref{magmag}(b) for $(\alpha, \delta)=(-0.39, 0.8)$ and $N=96$ with open boundary condition. It is clear that the magnetization starts from nonzero value less than $M_{\rm s}/3$ at $H=0$. The magnetization profile in this state  calculated by  the DMRG method is shown in  Fig. \ref{lm}(b) for the same set of parameters. As in the case of  Yoshikawa and Miyashita\cite{ym}, the magnetization profile  has the incommensurate modulation. 
\begin{figure}
\centerline{\includegraphics[width=20mm]{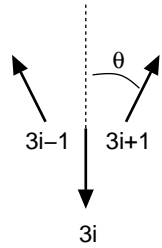}}
\caption{Classical noncollinear spin configuration.}
\label{ncol}
\end{figure}

The presence of the noncollinear ferrimagnetic state can be also understood within the classical picture. If we assume the noncollinear spin configuration depicted in Fig. \ref{ncol} and minimize the classical energy calculated using the Hamiltonian (\ref{ham}) with respect to the angle $\theta$, we find the nonzero solution of $\theta$ for $\delta > -1-3\alpha$ corresponding to the noncollinear ferrimagnetic phase. For $\delta < -1-3\alpha$, we find $\theta=0$ corresponding to the Lieb-Mattis type ferrimagnetism. However, the observed incommensurate ferrimagnetic spin profile cannot be understood within this classical picture. We expect this phenomenon is essentially due to the combined effect of quantum fluctuation and frustration.

Finally, in the narrow region between the ferrimagnetic phase and the spontaneously dimerized phase, another  Tomonaga-Luttinger liquid phase is found. Considering the difference in the spin structures of spontaneously dimerized phase and noncollinear ferrimagnetic phase, it is reasonable to expect an intermediate critical phase between these two phases. However, it still possible that this phase does not survive in the thermodynamic limit because of the limitation of the system size and ambiguity in the exterpolation procedure.
\section{Summary} 

The phase diagram of the  $S=1/2$ frustrated Heisenberg chains with period 3 exchange modulation is determined by analysing the exact numerical diagonalization data. In addition to the Tomonaga-Luttinger liquid phase, dimer phase, the Lieb-Mattis type ferrimagnetic phase and noncollinear ferrimagnetic phase are found. Physical interpretation of the  phase diagram based on the perturbational argument and classical picture is given. Typical magnetization curve in each phase is preented. It is shown that the magnetization profile has incommensurate modulation in noncollinear ferrimagnetic phase. This feature the similarity to the model investigated by Yoshikawa and Miyashita\cite{ym} which has also frustration and noncollinear ferrimagnetism. Therefore we may regard the incemmensurate spin profile as a characteristic of the noncollinear quantum ferrimagnetism induced by frustration.

Another Tomonaga-Luttinger liquid phase is found in the narrow region between  dimer phase and noncollinear ferrimagnetic phase.  Further investigation of the nature of this phase is left for future studies.
  
The author thanks S. Miyashita for enlightening comments and discussion. This work is partly supported by a Grant-in-Aid for Scientific Research from the Ministry of Education, Culture, Sports, Science and Technology, Japan.  The numerical diagonalization program is based on the package TITPACK ver.2 coded by H. Nishimori. The numerical computation in this work has been carried out using the facilities of the Supercomputer Center, Institute for Solid State Physics, University of Tokyo and the Information Processing Center, Saitama University.


\section*{References}
\begin{thebibliography}{10}
\bibitem{mg} Majumdar C K  and Ghosh D K :  1969 {\it J. Math. Phys.} {\bf 10} 1399.
\bibitem{oku1} Okunishi K  and Tonegawa T : 2003 {\it J. Phys. Soc. Jpn.} {\bf 72}  479.
\bibitem{oku2} Okunishi K and  Tonegawa T : 2003 {\it Phys. Rev.} {\bf B68}  224422.
\bibitem{tone} Tonegawa T, Okamoto K, Okunishi K, Nomura K and Kaburagi M : 2004 {\it Physica} {\bf B 346-347}  50.
\bibitem{ha} Hida K and Affleck I : 2005 {\it J. Phys. Soc. Jpn.} {\bf 74}  1849.
\bibitem{ym} Yoshikawa S and Miyashita S :  2005 {\it J. Phys. Soc.Jpn.} {\bf 74} Suppl. 71.
\bibitem{no} Nomura K and  Okamoto K : 1993 {\it J. Phys. Soc. Jpn} {\bf 62}  1123.
\end{thebibliography}
\end{document}